\documentclass{article}

\PassOptionsToPackage{numbers, sort&compress}{natbib}
\usepackage[preprint]{neurips_2025}
\usepackage{float} 

\usepackage[utf8]{inputenc} 
\usepackage[T1]{fontenc}    
\usepackage{hyperref}       
\usepackage{url}            
\usepackage{booktabs}       
\usepackage{amsfonts}       
\usepackage{nicefrac}       
\usepackage{microtype}      
\usepackage{xcolor}         
\usepackage{graphicx}
\usepackage{subcaption}
\usepackage{amsmath}
\usepackage{amssymb}
\usepackage{natbib}

\title{Dual‐Task Graph Neural Network for Joint Seizure Onset Zone Localization and Outcome Prediction using Stereo EEG 
}

\author{%
  Syeda Abeera Amir$^1$, Artur Agaronyan $^1$, William Gaillard $^{1,2}$, \\ \textbf{Chima Oluigbo $^{1,2}$, Syed Muhammad Anwar $^{1,2}$} 
    \\ \\
  {$^1$} Childrens National Hospital, Washington DC, USA\\
  {$^2$} George Washington University, Washington DC, USA \\
  \\
}

\begin{document}

\maketitle

\begin{abstract}
 Accurately localizing the brain regions that triggers seizures and predicting whether a patient will be seizure-free after surgery are vital for surgical planning and patient management in drug-resistant epilepsy. Stereo-electroencephalography (sEEG) delivers high-fidelity intracranial recordings that enable clinicians to precisely locate epileptogenic networks. However, the clinical identification is subjective and dependent on the expertise of the clinical team. Data driven approaches in this domain are sparse, despite the fact that sEEG offers high temporal-fidelity related to seizure dynamics that can be leveraged using graph structures ideal for imitating brain networks. In this study, we introduce a dual‐task graph‐neural network (GNN) framework that operates on windowed sEEG recordings to jointly predict seizure‐freedom outcomes and identify seizure‐onset‐zone (SOZ) channels. We assemble non-overlapping 10 second windows from 51 clinical seizures spread across 20 pediatric patients, with sEEG data annotated by clinical experts. For each temporal window we construct a functional connectivity graph via thresholded Pearson correlations and extract rich node features (spectral, statistical, wavelet, Hjorth and local graph features), alongside six global graph descriptors. We optimize a combined cross‐entropy loss with a tunable task‐weight, and select model hyper-parameters via Optuna. Under window‐level 10-fold cross‐validation, the model achieves a mean graph‐level accuracy of 89.31 ± 0.0976 {\%} for seizure‐freedom prediction and a node‐level SOZ localization accuracy of 94.72 ± 0.0041 {\%}. For the best performing model, we ran additive and leave-one-out ablation studies to explore feature importance for graph and node-level accuracy. 
 Our results demonstrate that jointly modeling global outcomes and local epileptogenic foci in connectivity graphs can significantly enhance prediction accuracy and interpretability, paving the way for AI‐assisted, network‐informed epilepsy surgery planning. 
\end{abstract}

\section{Introduction}

Epilepsy is one of the most prevalent neurological disorders globally, affecting over 70 million individuals across all demographics. It is characterized by uncontrolled electric discharges in a group of neurons, which result from abnormal electrical activity in the brain \cite{adeli_analysis_2003}. 
The condition often interferes with daily life, impacting cognitive function, emotional health, and social interactions \cite{milligan_epilepsy_2021, manford_recent_2017, operto_epilepsy_2023}. While anti-seizure medications are effective in managing symptoms in many cases, roughly 30${\%}$ of individuals with epilepsy suffer from drug-resistant epilepsy (DRE). In these circumstances, seizures continue despite administering pharmacological intervention \cite{miller_decision-making_2022, dwivedi_surgery_2017}. 

For individuals with DRE, epilepsy surgery is one of the most effective long-term solutions, with the potential to significantly reduce seizure frequency or even completely eliminate the occurrence of seizures. However, surgical success is heavily dependent on accurate identification of the seizure-onset zone (SOZ), the brain region primarily responsible for initiating seizures \cite{miller_decision-making_2022}. 
For cases where lesions are not identified using magnetic resonance imaging, methods for assessing seizure activity involve EEG-based localization using subdural grids and intracranial electrodes. In recent years, pioneering techniques incorporate high quality stereo electroencephalography (sEEG) as an alternative to time-consuming intracranial recordings that used grids, which may take days to weeks \cite{jiang_interictal_2022}. 
While these techniques provide a large  amount of precise and high temporal resolution data, effectively analyzing the vast amounts of generated information is an intensive, time consuming, and manual process that only specialized epileptologists can perform. 

Neuroscience and epilepsy research now recognize seizures as a network-driven phenomena rather than isolated focal events. Seizures are understood as disruptions in complex brain connectivity rather than originating purely from a single site \cite{lee_association_2024, niemeyer_seizure_2025}. This paradigm shift has fueled interest in utilizing computational models that quantify seizure spread and predict intervention outcomes with greater accuracy than would be possible with manual methods \cite{lam2024self, mutersbaugh2023epileptic, usman_epileptic_2021}. Among the most promising developments in this field is the application of graph neural networks (GNNs), which represent brain activity as a structured network rather than as discrete signals \cite{leng2025self, agaronyan2025graph, awais_graphical_2024}. GNNs differ from conventional signal-processing methods in their ability to capture dynamic relationships between brain regions. By treating electrodes as nodes and functional interactions as edges, GNNs provide a comprehensive opportunity for biologically informed model analyses of seizure propagation \cite{liu_comprehensive_2024}. This enables simultaneous analysis of localized SOZ activity and broader network dynamics, painting a more comprehensive picture of seizure propagation and genesis. Traditional methods of surgical planning rely on expert judgment informed by visual inspection of intracranial EEG such as sEEG recordings \cite{miller_decision-making_2022, dwivedi_surgery_2017}. This approach, while effective in many cases, is limited by human interpretation, potential bias, and inconsistencies in data analysis. Computational models powered by GNNs provide an alternative data-driven approach that can support clinicians by offering quantifiable and repeatable predictions as well as provide pioneering insights into seizure dynamics. By integrating machine learning techniques with clinical expertise in the form of training data, epilepsy surgery planning can transition from a subjective practice to a more precise, personalized approach tailored to each patient's unique seizure profile.  

By harnessing GNNs to model seizure propagation and intervention effects, we aim to provide clinicians with a powerful tool to model epileptic networks, giving patients a greater chance of achieving long-term seizure control and a better quality of life. Our proposed dual-task GNN framework builds upon this foundation by integrating seizure outcome prediction and SOZ localization into a single computational model. Previous studies have typically approached these tasks independently, but our method recognizes their inherent interdependence \cite{ramakrishnan_localization-related_2025, guo_seizure_2024}. By jointly optimizing both objectives, we take advantage of overlapping temporal and spatial seizure characteristics to enhance prediction accuracy and model applicability. The practical implications of our proposed framework could significantly improve clinical decision-making. 

\textbf{Our Contributions:} 
Our work provides the following major contributions:
\begin{itemize}
    \item {We propose a novel Dual-Task GNN for jointly predicting both the seizure onset zone and the seizure freedom outcomes for pediatric patients disgnosed with drug-resistant epilepsy.} 
     \item{We have curated a unique dataset from pediatric patients undergoing sEEG, featuring electrode placements in deep subcortical regions (such as thalamus). This dataset addresses a significant gap in existing research, which has predominantly focused on adult populations and cortical recordings.}
    \item{Our extensive analysis include features from multiple domains and we show how graph related features are significant in understanding the brain connectivity dynamics during seizure propagation. }
\end{itemize}

\subsection*{Related Works}

GNNs have been explored in recent years for the prediction of seizures and the classification of outcomes in multiple modalities of brain data. Li et al. (2022) proposed a dynamic graph-generative model for scalp EEG, achieving 91\% precision in the classification of multiclass seizure types \cite{li_graph-generative_2022}. Xiang et al. (2025) used a spatiotemporal GNN focused on synchronization in the CHB-MIT dataset, reporting 98\% precision and sensitivity in the prediction of early seizure \cite{xiang_synchronization-based_2025}. Grattarola et al. (2022) proposed an attention-based GNN using intracranial EEG (iEEG) recordings to localize SOZs by highlighting influential nodes in the brain graph \cite{grattarola_seizure_2022}. Similarly, Nejedly et al. (2025) introduced a GNN-based system that integrated interictal stereo-EEG (sEEG), electrode coordinates, and MRI features to automate SOZ localization in epilepsy surgery planning \cite{nejedly_leveraging_2025}. Their model, trained on data from 75 patients, achieved an area under the ROC curve of approximately 0.89, outperforming conventional neural networks. These studies demonstrate the efficacy of GNN in capturing spatial and temporal aspects of epileptogenic networks and reinforce the utility of using functional connectivity based on sEEG in clinical decision-making. 

However, the current methods either focus on identifying the seizures or only localize the SOZ, despite the inherent overlap in SOZ idenitifcation and seizure freedom outcomes. Since a precise SOZ localization has the potential to lead to better post-surgical outcomes. As research in epileptic surgical planning continues to evolve, incorporating pioneering computational models such as ours has the potential to significantly enhance how seizures are understood, mapped, and ultimately prevented.  

\section{Methods}
Our proposed dual-task GNN architecture is presented in Figure 1, where we incorporate both node- and graph-level features extracted from sEEG data creating representative graph structures. 
\begin{figure}
    \centering
    \includegraphics[width=0.8\linewidth]{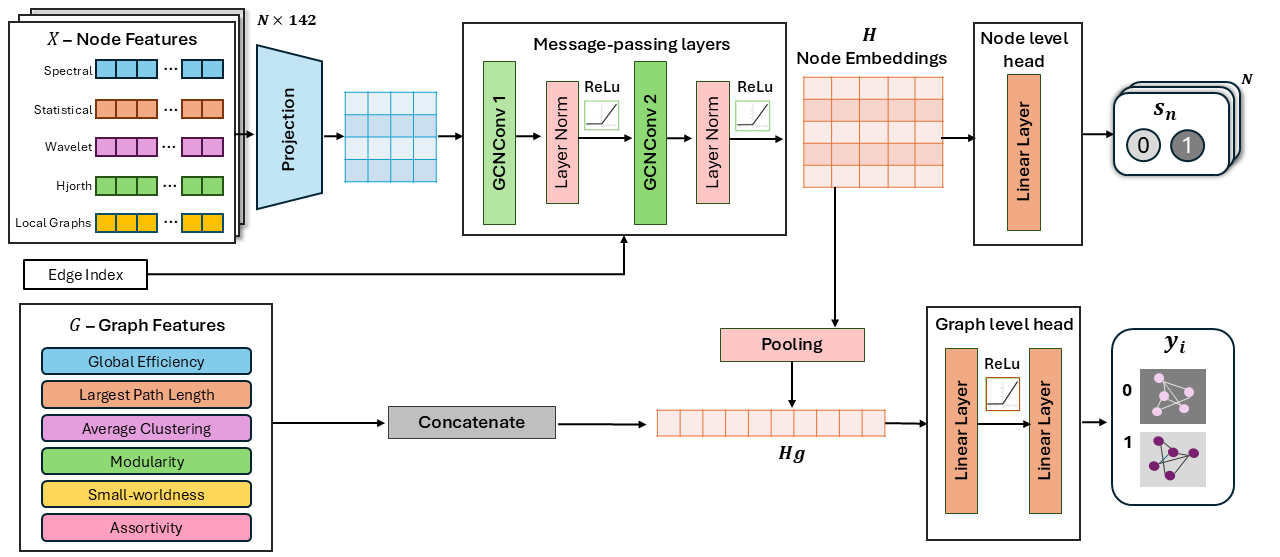}
    
    \caption{Our proposed Dual-Task Graph Neural Network model architecture. The diagram describes steps for analyzing sEEG input with $N$ channels for two outputs 1) $N$ node-level predictions, one for each node $n$, and 2) a single graph-level seizure freedom prediction for each sample $i$ with two classes 0: no worthwhile improvement, 1: seizure freedom.} 
    \label{fig:block-diagram}
\end{figure}
\subsection{Data Curation}
This study includes the sEEG data acquired from 20 pediatric patients at our clinical center diagnosed with pharmacoresistant epilepsy under an approved IRB protocol. A sample patient MRI with overlaid wires and contacts, and a post-op CT scan showing the contacts placement are shown in Figure 2a, and 2b respectively. Each wire can have up to 10 or more contacts, and the wire placement is decided based on clinical presentation. We have 10 patients who had post-operative seizure freedom following ablative surgery and 10 without seizure freedom. The seizure outcomes were recorded at 6 months follow up post-surgery and determined according to the Engel scale \cite{engel_practice_2003}. All data were obtained in routine clinical care. Relevant clinical information such as demographics, semiology, etiology, preoperative testing, results in noninvasive epilepsy investigations, details of sEEG procedure, definitive epilepsy surgery after sEEG, and seizure outcome at the last follow-up were extracted from patient records. A total of 51 seizures were considered (2-3 seizures per patient). The SOZ channels for each patient were identified and curated based on labeling from the clinical team involved in planning and performing the surgeries. The labels are characterized as either (1): involved in the seizure onset; or (0): not involved. The seizure outcomes were classified as (1) if the patient’s Engel score was in class I (seizure freedom); and (0) if it was in class II or below (no worthwhile improvement). 

\begin{figure}[htbp]
    \centering
    \begin{subfigure}[b]{0.35\linewidth}
        \centering
        \includegraphics[width=\linewidth]{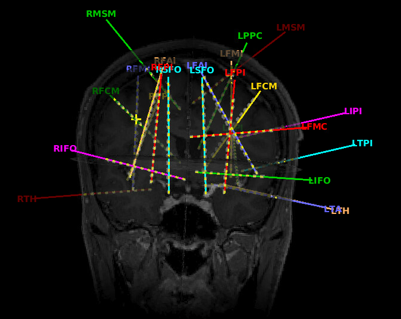}
        \label{fig:subfig-seeg-color}
    \end{subfigure}
    \hfill  
    \begin{subfigure}[b]{0.35\linewidth}
        \includegraphics[width=\linewidth]{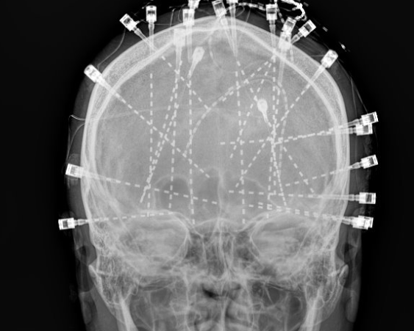}
        \label{fig:subfig-seeg-CT}
        \end{subfigure}
        
    \caption{Demonstration of (a) sEEG electrode placement within the brain and (b) post-op Computed Tomography scan showing sEEG contacts in the brain. 
    }
    \label{fig:seeg}
\end{figure}

\subsection{Data preprocessing and feature extraction}

To capture evolving network dynamics, each patient’s continuous sEEG recording was segmented into non-overlapping windows of length $W=10 sec$ (1387 samples at 512 Hz). Each window inherited the binary outcome label of its patient (free from seizures vs. non-free from seizures) and its SOZ mask per channel. This yielded on average ~70 windows per patient, augmenting the data for each seizure. Further, our proposed pipeline extracts a comprehensive set of features that capture both the temporal and spatial characteristics of sEEG signals, performed at two levels to emulate local- and network-wide descriptors of neuronal activity in the brain. For each window, node-level and graph-level features were calculated \cite{steuer2008global}. 

\paragraph{a) Node-level features}
are local to each channel in an sEEG temporal window. For each channel, we computed multiple feature types that collectively capture different, and complementary, information. The complete feature set for each channel comprises 142 dimensions that collectively capture spectral (PSD), statistical (distribution moments), temporal (Hjorth parameters, Wavelet energies), and network characteristics (local graph features).

\paragraph{Spectral features} are represented using the power spectral density (PSD) \cite{youngworth2005overview}. A sampling frequency of 128 Hz was used to downsample the features and remove redundancies. The features were calculated using $P_{i,b} = \frac{1}{T}\sum_{\omega\in b} S_{i}(\omega),$
where $P_{i,b}$ is the spectral power of channel $i$ in frequency band $b$ ($b {\in} {\delta},{\theta},{\alpha},{\beta},{\gamma}$), $T$ is the length of the time window in samples. $S_{i}(\omega)$ is the power spectral density of channel $i$ at frequency $\omega$.

\paragraph{Statistical moments} included mean ($\mu_i$), standard deviation ($\sigma^{2}_{i}$), skewness (${skew}_i$), and kurtosis (${kurt}_i$) of the zero-mean signal $x_i(t)$ for each channel $i$.

\paragraph{Hjorth parameters}\cite{oh2014novel} were used to capture the statistical properties in the time domain of the raw sEEG signal $x_i(t)$ from each channel $i$ including:
\begin{align*}    
    \mathrm{Activity}_i = \sigma_i^2, \quad
    \mathrm{Mobility}_i = \sqrt{\frac{\mathrm{Var}(\Delta x_i)}{\mathrm{Var}(x_i)}}, \quad and \quad
    \mathrm{Complexity}_i = \frac{\sqrt{\mathrm{Var}(\Delta^2 x_i)/\mathrm{Var}(\Delta x_i)}}{\mathrm{Mobility}_i}
\end{align*}
\paragraph{Wavelet energies}\cite{pathak2009wavelet} were computed using Wavelet decomposition. This enabled the analysis of signals at multiple time-frequency resolutions, capturing both transient and oscillatory components. The energy feature is calculated using $E_{i,k} = \sum_{t=1}^{T} c_{i,k}(t)^2,\quad k=1,\dots,4,$,
where $c_{i,k}(t)$ represents the wavelet coefficients computed from $x_i(t)$ at decomposition level $k$ and time point $t$.

\paragraph{Local graph features} were used to capture the network properties of individual channels within the broader brain network, we compute the following local graph metrics,
\begin{align*}
    \text{Strength: } s_i
    &= \sum_{j}A_{ij}, \quad and 
    &\text{Clustering: } C_i
    &= \frac{\sum_{j,k}A_{ij}A_{jk}A_{ki}}{\sum_{j,k}A_{ij}A_{ik}},
\end{align*}
where $N$ is the total number of channels (nodes) in the network and $A_{ij}$ is the adjacency matrix element representing the functional connectivity between channels $i$ and $j$.

Combining these node level features resulted in a feature vector $\mathbf{x}_i \in \mathbb{R}^F (F = 142)$. Concatenating across all $N$ channels yields $\mathbf{X} \in \mathbb{R}^{N{\times}F}$ for a temporal window $t$. This multi-dimensional feature representation enables our model to learn complex patterns associated with seizure onset, capturing both channel-specific activities and inter-channel dynamics essential for accurate SOZ localization.

\paragraph{b) Graph-level features}
In addition to node-level features, we extracted network-wide metrics that captured global topological properties of the sEEG signal during seizure activity. For each temporal window of length $T$, we first compute the zero-lag Pearson correlation matrix $|A_{ij}|$ from the raw signals $x_i(t)$ and $x_j(t)$. To focus on the most significant connections, we apply a threshold $\tau = 0.3$ to obtain a sparse weighted adjacency matrix $A_{ij}$ used for all subsequent graph calculations. From this thresholded matrix, we extract six global network summaries.

\textbf{Global efficiency} represented the measure of how quickly information can travel across the network and is the average of the inverse shortest path lengths between all node pairs calculated using \(E_{\mathrm{glob}} = \frac{1}{N(N-1)}\sum_{i\neq j}\frac{1}{d_{ij}}\), where \(N\) is the number of nodes and $d_{ij}$ is the shortest path length between nodes $i$ and $j$ in the thresholded network.

\textbf{Characteristic path length} represented the mean shortest path length within the network's largest connected component $C$ calculated using \(L = \frac{1}{N_c(N_c-1)}\sum_{i\neq j\in C}d_{ij}\), where $N_c$ is the number of nodes in component $C$.

\textbf{Average clustering} represented the mean of the clustering coefficient of each node $C_i$ in the network calculated using \( \bar C = \frac{1}{N}\sum_{i}C_i\), where $C_i$ is the clustering coefficient of node $i$, as defined earlier.  

\textbf{Modularity} represented the measure of the strength of division into modules (communities) via greedy algorithm community detection calculated using \( Q = \frac{1}{2m}\sum_{i,j}\Bigl[A_{ij}-\frac{s_i s_j}{2m}\Bigr]\delta(c_i,c_j)\), where $m$ is the total weight of all edges in the network and $s_i$ is the strength of node $i$, as defined earlier. 

\textbf{Small-World index} represented a ratio comparing clustering and path length to a random graph baseline calculated using \( \sigma = \frac{\bar C / C_{\mathrm{rand}}}{L / L_{\mathrm{rand}}}\). Values $\sigma >1$ denotes “small-world” networks.

\textbf{Assortativity} represented the property that quantifies the preference for nodes to attach to others that are similar (or dissimilar) in terms of degree calculated using $ r = \frac{\sum_{i,j}(A_{ij}-\tfrac{s_i s_j}{2m})\,s_i s_j}
             {\sum_{i,j}(s_i\delta_{ij}-\tfrac{s_i s_j}{2m})\,s_i s_j}$, where $s_i$ (strength/degree) and $m$ (total weight) are as defined earlier.

These six graph-level features collectively characterize the global topological properties of the brain network during each temporal window $T$. They form a graph feature vector $\mathbf{G} \in \mathbb{R}^6$ that complements the node-level features ($\mathbf{X} \in \mathbb{R}^{N \times F}$) in our analysis pipeline.

\subsection{Model development}

Following preprocessing, the segments were structured into graph representations, where each sEEG channel served as a node. Temporal edges were constructed between nodes on correlation patterns above a threshold value ($\tau$ = 0.3). The node features $X$, graph features $G$, and edges index $E$ for each temporal window were used as inputs for the model. The model architecture is shown in Figure {\ref{fig:block-diagram}}. It consists of two message-passing layers of graph convolutional operators, with additional linear layers for projection and classification. For node-level classification, node embeddings $H{\in}R^{N{\times}D}$, from node features were used to classify the labels for each node (SOZ or no SOZ), resulting in $N$ predictions per window, where $N$ is the number of channels for that segment. For graph-level classification, node embeddings $H$ across all nodes were pooled using $\mathbf{h}_{\text{pool}} = \text{MEAN\_POOL}(\mathbf{H}^{(2)}, \mathbf{batch}) \in \mathbb{R}^{D}$ and then concatenated with the graph features $G$ using $\mathbf{h}_{\text{combined}} = [\mathbf{h}_{\text{pool}} \parallel \mathbf{g}] \in \mathbb{R}^{D+6}$,
where $\mathbf{batch}$ is a batch assignment vector. The concatenated vector $\mathbf{h}_{\text{combined}}$ is then processed through a two-layer multi layer perceptron to generate $H_g$. The resulting vector $H_g$ was then passed through a two-layer graph classification head, which output a single prediction per window: seizure freedom (class 1) or no seizure freedom (class 0).

We employed a weighted dual-task loss function that balances node-level and graph-level classification heads represented as, 
\begin{align*}
\mathcal{L}_{\text{total}} = (1-\alpha) \cdot \mathcal{L}_{\text{graph}} + \alpha \cdot \mathcal{L}_{\text{node}},
\end{align*}
where $\alpha \in [0,1]$ is a hyperparameter that controls the contribution of each task. For both tasks, we use cross-entropy loss, and to address class imbalance, particularly in the node classification task where SOZ nodes are the minority, we incorporated class weights in the loss function.

\section{Experiments and Results}

We conducted several experiments to explore the model’s performance across different combinations of features and settings. For our base model, we performed a 10-fold seizure-level cross-validation, training on 60{\%} of the sEEG segments, while splitting the rest of the 40{\%} for validation and testing. In comparison, we ran a patient-wise analysis where we employed a Leave-One-Out cross validation (LOOCV) approach, sequentially training the dual-task GNN model on all patient graphs except for one patient, which was used as the test case. Hyperparameter tuning with 100 trials and 30 epochs per trial was done using Optuna \cite{akiba2019optuna}. Training and validation was conducted on Nvidia RTX A5000 GPU, with a total runtime of approximately 1 hour. 

We report accuracy, precision, recall and F1 score for both graph-level (outcome prediction) and node-level (SOZ localization) classification on the held-out test windows. Table 1 shows a summary of results for seizure-wise and patient-wise analysis. 
\begin{table}[htbp]
    \centering
    
    \caption{Performance metrics for seizure-wise and patient-wise analysis.}
    
    \label{tab:seizure_detection}
    \resizebox{0.6\textwidth}{!}{
    \begin{tabular}{lcccc}
        \toprule
        & \multicolumn{2}{c}{\textbf{Seizure-wise (Base model)}} & \multicolumn{2}{c}{\textbf{Patient-wise}} \\
        \cmidrule(lr){2-3} \cmidrule(lr){4-5}
        \textbf{Metric} & \textbf{Graph-level} & \textbf{Node-level} & \textbf{Graph-level} & \textbf{Node-level} \\
        \midrule
        \textbf{Accuracy} & 0.89 $\pm$ 0.097 & 0.94 $\pm$ 0.004 & 0.87 $\pm$ 0.252 & 0.88 $\pm$ 0.204 \\
        \textbf{Precision} & 0.90 $\pm$ 0.085 & 0.90 $\pm$ 0.006 & 0.84 $\pm$ 0.206 & 0.84 $\pm$ 0.206 \\
        \textbf{Recall} & 0.89 $\pm$ 0.097 & 0.94 $\pm$ 0.004 & 0.88 $\pm$ 0.204 & 0.88 $\pm$ 0.204 \\
        \textbf{F1} & 0.88 $\pm$ 0.010 & 0.92 $\pm$ 0.004 & 0.86 $\pm$ 0.206 & 0.86 $\pm$ 0.206 \\
        \bottomrule
    \end{tabular}}
\end{table}

We also assessed the dual-task GNN model’s performance across a combination of features. The base model includes 5 types of node features $X$, and 6 global graph descriptors combined in vector $G$. We tested the model’s performance without the graph features and ran an additive and leave-one-out ablation study for the node features. Table 2 highlights the model’s performance without using $G$ as an input and training on only node features, $X$. There is a significant drop in the node-level accuracy, going from 0.94 to 0.86 when input feature $G$ is omitted, while the graph-level accuracy only has a slight decrease, going from 0.89 to 0.88. 

\begin{table}[htbp]
    
    \centering
    \caption{Performance metrics without graph features.}
    
    \label{tab:without_graph_features}
    \resizebox{0.35\textwidth}{!}{
    \begin{tabular}{lcc}
        \toprule
        & \multicolumn{2}{c}{\textbf{Without Graph Features}} \\
        \cmidrule(lr){2-3}
        \textbf{Metric} & \textbf{Graph-level} & \textbf{Node-level} \\
        \midrule
        \textbf{Accuracy} & 0.88 $\pm$ 0.13 & 0.86 $\pm$ 0.26 \\
        \textbf{Precision} & 0.89 $\pm$ 0.12 & 0.91 $\pm$ 0.01 \\
        \textbf{Recall} & 0.88 $\pm$ 0.13 & 0.86 $\pm$ 0.26 \\
        \textbf{F1} & 0.86 $\pm$ 0.16 & 0.83 $\pm$ 0.26 \\
        \bottomrule
    \end{tabular}}
\end{table}

\subsection{Network Analysis}

To investigate the spatio-temporal dynamics of neural synchronization during seizure onset, we implemented a network analysis framework. This approach enables visualization and quantification of functional connectivity patterns between electrodes, with special emphasis on the seizure onset zone. Functional connectivity was quantified using Phase Locking Value (PLV) \cite{lachaux_measuring_1999}, a widely utilized measure of phase synchronization between two signals. For each pair of electrodes, PLV was calculated as: 
\begin{align*}
\text{PLV}{x,y} = \left| \frac{1}{N} \sum{t=1}^{N} e^{i(\phi_x(t) - \phi_y(t))} \right|,
\end{align*}
where $(\phi_x(t))$ and $(\phi_y(t))$ represent the instantaneous phases of signals $(x)$ and $(y)$ at time $(t)$, derived through the Hilbert transform, and $N$ is the number of time points in the analysis window. PLV values range from 0 to 1, where 0 indicates no phase synchronization and 1 indicates perfect phase locking between the signals. Instantaneous phase was extracted by applying the Hilbert transform to each electrode's time series using $phi_x(t) = \arg(x(t) + i\mathcal{H}{x(t)}$, 
where $(\mathcal{H}{x(t)})$ denotes the Hilbert transform of signal $(x(t))$.

\begin{figure}[!t]
    \centering
    \begin{subfigure}{\linewidth}
        \centering
        \includegraphics[width=0.8\linewidth]{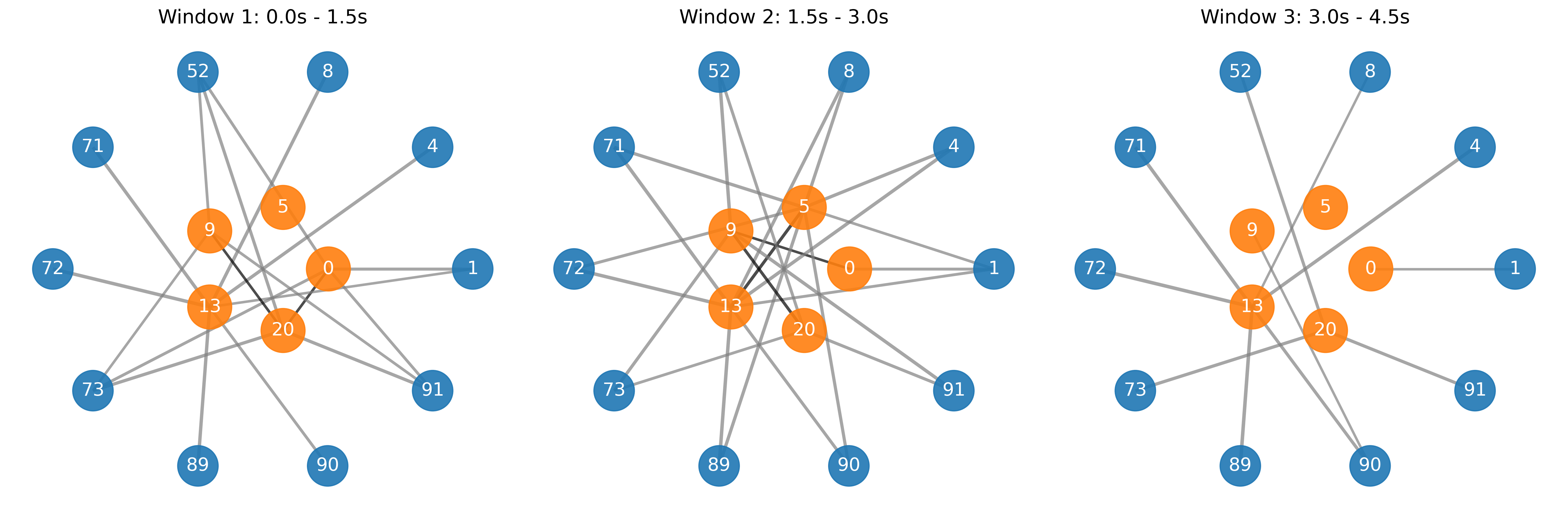}
        \caption{Sample from class 0: no seizure freedom}
        \label{fig:0-network}
    \end{subfigure}
    
    \vspace{1em}  
    
    \begin{subfigure}{\linewidth}
        \centering
        \includegraphics[width=0.8\linewidth]{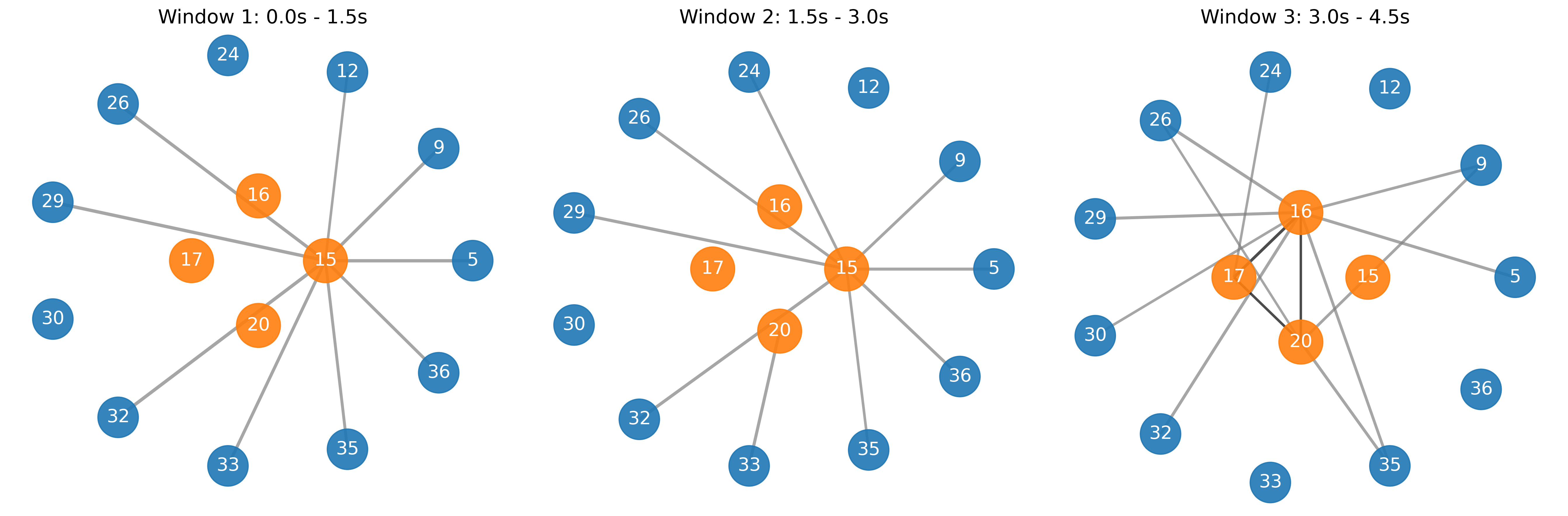}
        \caption{Sample from class 1: seizure freedom}
        \label{fig:1-network}
    \end{subfigure}
    
    \caption{Network dynamics at seizure onset from samples (a) without seizure freedom, (b) with seizure freedom post-surgery. The figure shows all SOZ nodes for both patients and a subset of the most well-connected nodes to the SOZ nodes. Edges are calculated using phase locking values, thresholded at ${\theta} = 0.65$. The orange nodes represent SOZ nodes while the blue ones are non-SOZ nodes. The thickness of the edges highlights connection strength. The same subset of nodes is used in all three windows to demonstrate seizure dynamic in one network. The node numbers are assigned as replacement to the clinical names given to each of these sEEG contacts.}
    \label{fig:network-dynamics}
\end{figure}

\begin{table}[!t]
    \centering
    
    \caption{Network properties across time windows for PLV threshold $\theta = 0.65$.}
    
    \label{tab:network_properties}
    \resizebox{0.6\textwidth}{!}{
    \begin{tabular}{lcccccc}
        \toprule
        & \multicolumn{3}{c}{\textbf{Class 0}} & \multicolumn{3}{c}{\textbf{Class 1}} \\
        \cmidrule(lr){2-4} \cmidrule(lr){5-7}
        \textbf{Metric} & \textbf{W1} & \textbf{W2} & \textbf{W3} & \textbf{W1} & \textbf{W2} & \textbf{W3} \\
        \midrule
        Density & 0.180 & 0.228 & 0.104 & 0.085 & 0.085 & 0.133 \\
        Average Clustering & 0.226 & 0.708 & 0.000 & 0.000 & 0.000 & 0.256 \\
        Average SOZ degree & 4.200 & 5.400 & 2.200 & 2.250 & 2.250 & 4.250 \\
        Average SOZ PLV & 0.712 & 0.781 & 0.000 & 0.000 & 0.000 & 0.710 \\
        \bottomrule
    \end{tabular}}
\end{table}
Figures {\ref{fig:0-network}} and {\ref{fig:1-network}} visualize the evolution of network dynamics at seizure onset of two random patients, one from each class. For each time window, we constructed weighted functional networks where the nodes represent individual electrodes, the edges represent the PLV between electrode pairs, included only if it exceeds a threshold $\theta$ (typically set to 0.65), and the SOZ nodes are highlighted to track their evolving connectivity patterns. To retain focus on seizure dynamics, only a small subset of high-degree nodes were included and only connections involving at least one SOZ node were visualized. 

For each time window graph $(w = 2 s)$, we report various network metrics to compare the network behavior of both samples, summarized in Table \ref{tab:network_properties}. In the peak density windows of both samples, class 0 (W2) has a higher density ($\rho_0 = 0.228, \rho_1 = 0.133$) and average clustering ($C_0 = 0.708, C_1 = 0.256$), compared to class 1 (W3). The SOZ nodes in class 0 are also more widely connected and synchronized, with the highest SOZ degree average of $5.4$ and SOZ PLV average of $0.781$, compared to $2.25$ and $0.71$, respectively, of class 1.    

\subsection{Ablation Studies}

\begin{figure}[!t]
    \centering
    \begin{subfigure}[b]{0.45\textwidth}
        \centering
        \includegraphics[width=\textwidth]{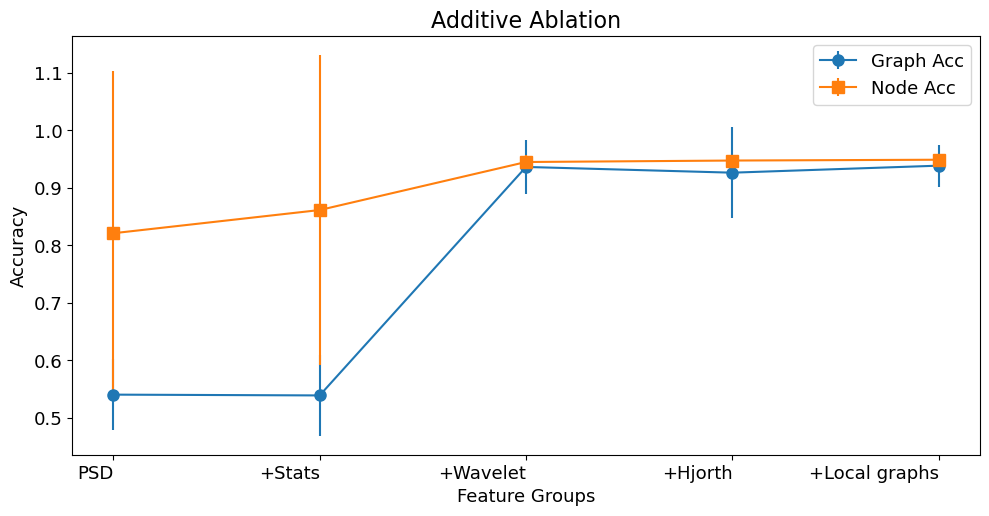}
        \caption{Additive ablation accuracies}
        \label{fig:subfig-additive}
    \end{subfigure}
    \hfill  
    \begin{subfigure}[b]{0.45\textwidth}
        \centering
        \includegraphics[width=\textwidth]{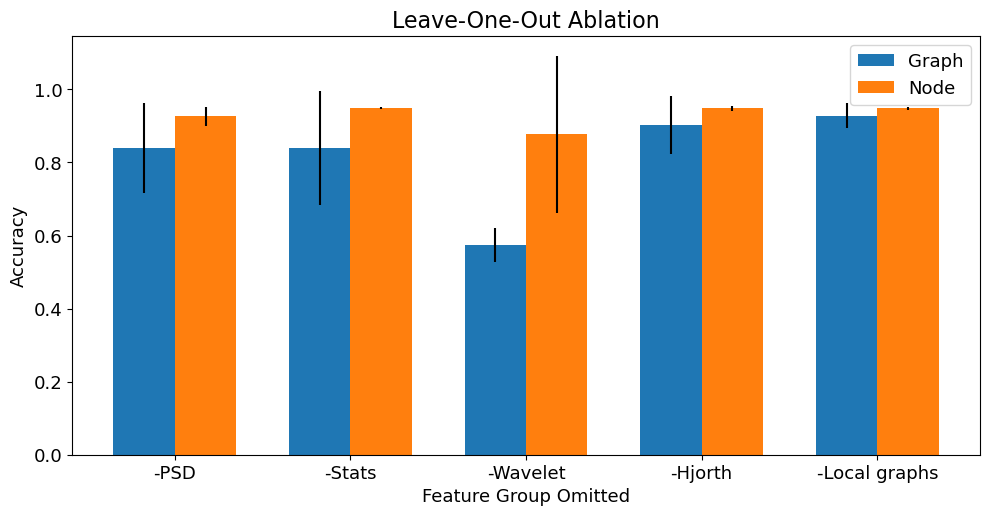}
        \caption{Leave-one-out ablation accuracies}
        \label{fig:subfig-LOO}
        \end{subfigure}
        
    \caption{Comparison of the dual-task GNN's performance for two types of ablation results. (a) shows changes in model's accuracy with sequential addition of features and (b) shows the changes in model's performance with the iterative exclusion of features.}
    \label{fig:both-results}
\end{figure}

To quantify the contribution of each feature group, we designed two complementary ablation experiments. 1) Additive: Start with spectral features alone, then incrementally add Statistical, Hjorth, Wavelet, and Local-graph features in that order. 2) Leave-one-out: Iteratively train using all feature groups except one (e.g. all minus PSD) to measure the performance drop when that group is removed. For each ablation setting, we retrained the full dual-task GNN (with the same hyperparameters selected via a preliminary Optuna sweep) and recorded test‐set accuracies. This yielded performance accuracy showing how each feature group improved or degraded graph and node classification when included or excluded, summarized in figures {\ref{fig:subfig-additive}} and {\ref{fig:subfig-LOO}}

\section{Discussion}

\begin{figure}[!t]
    \centering
    \includegraphics[width=0.35\linewidth]{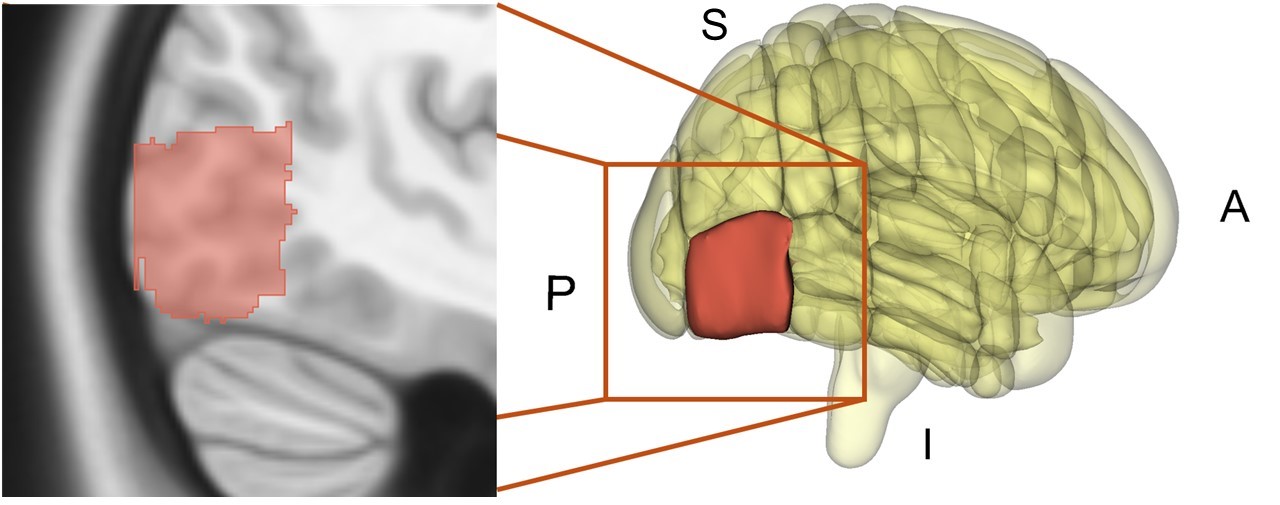}
    \caption{An example of the SOZ identified by our model representing the posterior insula in agreement with manual clinical analysis on a 3D parcellated brain (shown in red).}
    \label{fig:SOZ-viz}
\end{figure}

This study presents a dual-task GNN that jointly performs seizure outcome prediction and SOZ localization using sEEG data. Our results demonstrated that the model performs well in both seizure-wise and patient-wise cross-validation schemes, suggesting it is effective not only in intra-subject generalization but also in generalizing across patients; a significant benefit for clinical and surgical decision-making. These results show that our proposed GNN-based algorithm could inform clinical understanding of seizure propagation and network dynamics. Epileptic seizures are 
dynamic network phenomena involving distributed brain regions over time. Our additive ablation results further support this idea as graph-level accuracy increases only after introducing PSD followed by wavelets, then stays consistent after adding Hjorth and local graph features. This may be because PSD features do not incorporate global trends in connectivity, supporting the idea that seizures propagate in cooperation with many brain regions. Node-level accuracy remained consistent through the inclusion and exclusion of node features. Figure \ref{fig:SOZ-viz} demonstrates our model's high spatial accuracy when predicting the localized SOZ where contacts in the posterior insula are identified as the SOZ in agreement with the clinical findings.

Ablation analysis further revealed that excluding wavelets led to a significant reduction in graph accuracy. These features encapsulate frequency changes over time, suggesting that seizures have a global temporal complexity. However, node-level accuracy was affected the most by the exclusion of graph-features $G$, as shown in Table \ref{tab:without_graph_features}, dropping from 0.94 to 0.86. This could be because of the combined loss function that is propagated back in the model, and opens interesting avenues for further investigation.  

Our architecture provides a practical advantage by enabling simultaneous inference of patient-level outcomes and electrode-level SOZ localization for surgical planning. Precise localization enables neurosurgeons to identify optimal targets for resective surgery while minimizing the removal of non-epileptic brain tissue, preserving cognitive function and surgical risks. In addition, insights into global outcomes allow personalized surgical approaches, where the decision to perform focal resection, network disconnection, or neuro-modulation is guided by predicted patterns of seizure spread. 
Additionally, the network analysis shows that seizure onset exhibits distinct connectivity patterns, where patients with seizure freedom display lower synchronization in SOZ nodes compared to those without seizure freedom. Patients without seizure freedom exhibit higher average SOZ degree (5.4) and stronger synchronization, reflected in an average SOZ PLV of 0.781 during peak density windows, compared to seizure-free patients who show a reduced SOZ degree of 2.25 and lower synchronization with an average PLV of 0.710. Additionally, clustering coefficients demonstrate a clear contrast, where seizure-free patients maintain a lower average clustering (0.256) while non-seizure-free patients reach a much higher clustering value (0.708), suggesting a more interconnected and synchronized seizure onset pattern in the latter group.

Beyond surgical applications, our framework offers exciting possibilities for adaptive seizure management, including closed-loop neuromodulation and personalized pharmacological interventions. Understanding how seizures evolve in a patient-specific manner could enable real-time therapeutic adjustments, where stimulation is delivered dynamically to prevent seizure progression before it fully develops. Additionally, mapping seizure spread patterns could guide the development of targeted drug therapies, where medications are designed to specifically disrupt high-risk propagation pathways. As neuroscience increasingly intersects with machine learning, the integration of graph-based models into clinical workflows could improve not only pre-surgical evaluations but also non-invasive epilepsy treatment, offering a precision medicine approach to seizure control. 

\paragraph{Limitations:}
Despite its promising performance, the study has several limitations that warrant consideration. Our analysis was carried out on a relatively small cohort of 20 patients, resulting in high variability in our performance metrics, especially for the patient-wise analysis. The model tends to overfits on some patients while underfitting on others. Although we used segment-level cross-validation to mitigate overfitting, such a small sample size limits the statistical power of our findings. 
The patient group was drawn from a single center with similar clinical protocols and electrode implantation strategies; we do not have enough evidence to evaluate whether our dual-task GNN would perform equally well on data from other institutions and different age groups (e.g., pediatric vs. adult). 
We segmented the sEEG data into fixed 10s windows with no overlap. 
More adaptive or event-driven windowing, or model that designed to explicitly capture sequence information (e.g., temporal GNNs or transformers), might better leverage the data. Additionally, we built adjacency from zero-lag Pearson correlations, thresholded at fixed values $(\tau=0.3)$. Although correlation graphs are widely used, they capture only linear, stationary associations and ignore directionality and delays, features that might be crucial for seizure propagation. Nonlinear connectivity measures (e.g. mutual information, phase-locking value) or causality estimates (e.g. Granger causality, transfer entropy) could yield richer network representations but come at increased computational and methodological complexity that our small sample size might not be able to handle. 

\section{Conclusions}

This study demonstrate the power of GNNs in advancing data-driven epilepsy treatment and surgical planning. Our findings underline the importance of incorporating global connectivity features into predictive models, as seizures unfold through interactions across multiple brain regions rather than originating from isolated focal points. The robustness of our model across both seizure-wise and patient-wise analyses suggests that it can generalize well to diverse patient populations, marking an important step toward personalized epilepsy care. By demonstrating how GNN-powered analyses can reveal key trends in seizure dynamics, our work contributes to the broader effort of bringing data-driven solutions to epilepsy care.

\bibliographystyle{unsrtnat}
\bibliography{References}

\end{document}